\documentclass[pdflatex,sn-mathphys-num]{sn-jnl}

\usepackage{graphicx}%
\usepackage{multirow}%
\usepackage{amsmath,amssymb,amsfonts, bm}%
\usepackage{amsthm}%
\usepackage{mathrsfs}%
\usepackage[title]{appendix}%
\usepackage{xcolor}%
\usepackage{textcomp}%
\usepackage{manyfoot}%
\usepackage{booktabs}%
\usepackage{algorithm}%
\usepackage{algorithmicx}%
\usepackage{algpseudocode}%
\usepackage{listings}%

\begin{document}

\title[RiNALMo]{RiNALMo: General-Purpose RNA Language Models Can Generalize Well on Structure Prediction Tasks}


\author*[1]{\fnm{Rafael Josip} \sur{Penić}}\email{rafael-josip.penic@fer.unizg.hr}
\author*[2]{\fnm{Tin} \sur{Vlašić}}\email{tin\_vlasic@gis.a-star.edu.sg}
\author[3]{\fnm{Roland G.} \sur{Huber}}\email{rghuber@bii.a-star.edu.sg}
\author[2]{\fnm{Yue} \sur{Wan}}\email{wany@gis.a-star.edu.sg}
\author*[2]{\fnm{Mile} \sur{Šikić}}\email{mile\_sikic@gis.a-star.edu.sg}

\affil[1]{\orgdiv{Faculty of Electrical Engineering and Computing}, \orgname{University of Zagreb}, \orgaddress{\street{3 Unska Street}, \city{Zagreb}, \postcode{10000}, \country{Croatia}}}

\affil[2]{\orgdiv{Genome Institute of Singapore (GIS)}, \orgname{Agency for Science, Technology and Research (A*STAR)}, \orgaddress{\street{60 Biopolis Street, Genome}, \city{Singapore}, \postcode{138672}, \country{Republic of Singapore}}}

\affil[3]{\orgdiv{Bioinformatics Institute (BII)}, \orgname{Agency for Science, Technology and Research (A*STAR)}, \orgaddress{\street{30 Biopolis Street, Matrix}, \city{Singapore}, \postcode{138671}, \country{Republic of Singapore}}}


\abstract{While RNA has recently been recognized as an interesting small-molecule drug target, many challenges remain to be addressed before we take full advantage of it. This emphasizes the necessity to improve our understanding of its structures and functions. Over the years, sequencing technologies have produced an enormous amount of unlabeled RNA data, which hides a huge potential. Motivated by the successes of protein language models, we introduce RiboNucleic Acid Language Model (RiNALMo) to unveil the hidden code of RNA. RiNALMo is the largest RNA language model to date, with 650M parameters pre-trained on 36M non-coding RNA sequences from several databases. It can extract hidden knowledge and capture the underlying structure information implicitly embedded within the RNA sequences. RiNALMo achieves state-of-the-art results on several downstream tasks. Notably, we show that its generalization capabilities overcome the inability of other deep learning methods for secondary structure prediction to generalize on unseen RNA families.}




\maketitle

\section{Introduction}
\label{intro}
Large language models (LLMs) trained on massive text corpora have been performing remarkably on various natural language understanding and generation tasks \citep{devlin2018_bert, radford2018improving, raffel2020exploring, brown2020language, touvron2023llama, Chowdhery2023_palm}. In recent years, the exploration of language models (LMs) has gone beyond the domain of natural language processing (NLP), reaching into the realms of biology and its data. A vast amount of sequenced protein data provided a ground for training protein LMs, and since they have proven to be an extremely valuable asset in protein generative \citep{ferruz2022protgpt2, madani2023_ProtGen, nijkamp2023_progen2} and structure prediction tasks \citep{wu2022_omega_fold, Lin2023_esm_2}.

Most efforts in applying the ideas originally developed for NLP have been focused on proteins after the success of AlphaFold \cite{jumper2021alphafold2} in the prediction of protein structures. ESM-1b \cite{rives2021_esm_1b} was one of the first models that applied the self-supervised language modeling approaches to protein data. It was pre-trained on $250$M protein sequences and tested on several downstream tasks, including the secondary structure and tertiary contact prediction, where it achieved state-of-the-art results. Later on, several other protein LMs were proposed and tested on various downstream tasks \cite{heinzinger2019modeling, nambiar2020transforming, brandes2022proteinbert, Elnaggar2022_prott5}. Protein LMs play an important role in protein tertiary structure prediction. ESM-2 \cite{Lin2023_esm_2} and OmegaPLM \cite{wu2022_omega_fold} are examples of protein LMs that efficiently replace a multiple sequence alignment (MSA) step in deep learning (DL) methods for structure prediction.

RNAs play crucial roles in fundamental biological processes, including transcription, cell signaling, chromatin remodeling, and genome imprinting. Like proteins, RNAs have recently become an attractive drug target, whose function and interaction with other molecules are closely related to their structure \citep{childs2022targeting, garner2023contemporary}. However, much less attention has been given to applying LMs to RNA-related problems, partly because there is no such amount of available data and corresponding structures, and partly because similar problems tend to be more difficult than for proteins.

Currently, there are only two single-input-sequence RNA foundation models, RNA-FM \citep{chen2022_rna_fm} and Uni-RNA \citep{Wang2023_unirna}, that found applications in several structure and function prediction tasks. RNA-FM is a $100$M parameters Transformer encoder based on the original implementation by \citet{vaswani2017_attention} and trained exclusively on $23.7$M non-coding RNAs (ncRNAs) from the RNAcentral database \cite{rnacentral2021}. \citet{Wang2023_unirna} pre-trained an ensemble of LMs ranging from $25$M to $400$M parameters trained on a much larger dataset of $1$B sequences with the architecture analogous to the ESM protein LM \citep{rives2021_esm_1b} enhanced by several advanced techniques such as RoPE and fused layer norm. The authors pre-trained language models of different sizes and reported when the model parameters exceeded $400$M, the performance in downstream tasks reached a plateau with their architectures and datasets. Both foundation models used the standard BERT-style masked language modeling (MLM) pre-training task \citep{devlin2018_bert}. Unlike the Uni-RNA, RNA-FM is publicly available.

Besides the RNA-FM and Uni-RNA foundation models, several authors proposed LMs to solve a few specific downstream tasks. RNA-MSM \cite{zhang2024_msa_based_rna_lm}, an MSA-based BERT-style RNA LM, specialized in particular for secondary structure prediction, that instead of a single input sequence utilizes a set of homologous sequences. However, obtaining the MSAs is a very time-consuming procedure--it takes RNAcmap, a homology search tool used by RNA-MSM, on average nine hours to obtain an MSA for one RNA sequence of length $60$ \cite{zhang2024_msa_based_rna_lm}. \citet{Chen2023_splice_bert} proposed SpliceBERT, a BERT-style encoder pre-trained exclusively on more than $2$M precursor messenger RNA (pre-mRNA) sequences from different vertebrates for studying RNA splicing. SpliceBERT outperforms DNABERT \cite{ji2021_dna_bert}, an LM trained only on a human genome, both on human and non-human splice-site prediction tasks. It demonstrates better generalization capability of LMs pre-trained on multiple species. \citet{yang2022scbert} proposed single-cell BERT (scBERT), a BERT-style LM pre-trained on huge amounts of unlabeled single-cell RNA-seq data for cell type annotation. BigRNA \cite{celaj2023_bigrna} is an LM pre-trained on the genomes of $70$ individuals on a task to predict DNA-matched RNA-seq data. BigRNA accurately predicts tissue-specific RNA expression and the binding sites of proteins and microRNAs.

Motivated by the recent successes of protein LMs and the latest architectural improvements in LLMs, we propose RiNALMo, a novel RNA language model. We pre-trained RiNALMo on a set of carefully curated $36$M ncRNA sequences from the RNAcentral database augmented by several other RNA databases. RiNALMo is a $650$M parameters BERT-style Transformer encoder advanced by modern architectural techniques such as rotary positional embedding (RoPE) \cite{su2024_roformer}, SwiGLU activation function \cite{shazeer2020_swiglu} and FlashAttention-2 \cite{dao2023_flashattention2}. During pre-training, RiNALMo can extract hidden knowledge and capture the underlying structural information embedded within the sequences at the single-nucleotide level. Later, its output embeddings serve as a powerful sequence representation that improves the performance on various structural and functional RNA downstream tasks compared to other foundation models and state-of-the-art methods. In particular, RiNALMo shows remarkable generalization capability on secondary structure prediction of RNA families not encountered in the training dataset where other DL methods fail.

The main contributions of the paper are as follows:
\begin{itemize}[nolistsep]
    \item We propose RiNALMo, a $650$M parameters RNA LM, which is the largest RNA language model to date that can fully leverage the potential of a vast amount of public unannotated RNA sequences;
    \item We show that the generalization capability of RiNALMo can overcome the problem of other DL methods for secondary structure prediction to perform well on RNA families not seen in the training dataset;
    \item We conducted extensive experiments on several RNA structural and functional downstream tasks whose results show that RiNALMo outperforms other RNA LMs and DL methods on most datasets.
    \item We release the pre-trained and fine-tuned RiNALMo weights and scripts for fine-tuning the model for the downstream tasks.
\end{itemize}

\section{Results}
\label{sec:results}
\begin{figure*}[t]
    \centering
    \includegraphics[width=\textwidth]{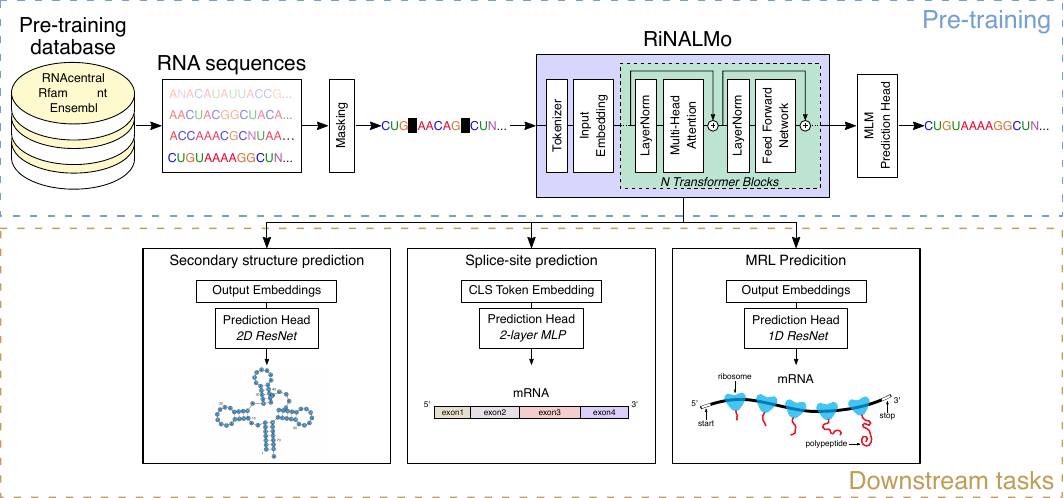}
    \caption{\textbf{RiNALMo pre-training and applications.} In the pre-training stage, RiNALMo is trained on unlabeled RNA sequences from several databases using masked language modeling (MLM). To corrupt the input sequence, we randomly mask $15\%$ of the tokens in the training sequence. Before being passed to the Transformer, an RNA sequence is tokenized and turned into a $1280$ dimension vector using a learned input embedding module. The language model comprises $33$ Transformer blocks. Each Transformer block consists of a multi-head attention and a feed-forward network. Once pre-trained, RiNALMo can be separately fine-tuned for various structural and functional downstream tasks in which its expressive output embeddings, utilized by the prediction heads, significantly improve the performance. In this work, we fine-tuned RiNALMo for secondary structure, multi-species splice-site and mean ribosome loading (MRL) prediction tasks.}
    \label{fig:rinalmo}
\end{figure*}

\bmhead{General-purpose RNA language model}
A schematic diagram of RiNALMo and its pre-training procedure and downstream tasks are shown in Fig. \ref{fig:rinalmo}.
Our LM is a Transformer encoder focused on understanding and unveiling the RNA code. At the heart of the model is the self-attention mechanism \cite{vaswani2017_attention}, which captures important local and global contextual information. We pre-trained RiNALMo using the MLM where we tasked the model to reconstruct corrupted unlabeled RNA sequences. In this paper, each nucleotide is a single token. To corrupt the input sequence, we randomly mask $15\%$ of the tokens in the training sequence. To reconstruct the masked tokens, RiNALMo's embeddings are utilized by the MLM prediction head whose outputs are used in the cross-entropy loss function. More technical details, pre-training details and ablation study are given in Methods and Supplementary materials.

Once pre-trained, RiNALMo's output embeddings can serve as a powerful sequence representation that has embedded structural and evolutionary information. First, its embeddings can be used for visualization and clustering analysis of RNA sequences. Second, such a representation can be used as an enriched input to structural and functional downstream tasks. We employed RiNALMo in a few tasks to assess its performance and generalization capabilities. Namely, we show how RiNALMo can improve and generalize well on secondary structure, multi-species splice-site and mean ribosome loading (MRL) prediction tasks. However, we anticipate it can be leveraged in many other tasks related to RNA structure and function. Particularly interesting would be the employment of RiNALMo in RNA tertiary structure prediction tasks where, motivated by the results from ESMFold \cite{Lin2023_esm_2} and OmegaFold \cite{wu2022_omega_fold}, we believe RiNALMo's embeddings can successfully replace the MSA.

\begin{figure*}[t]
    \centering
    \includegraphics[width=0.9\textwidth]{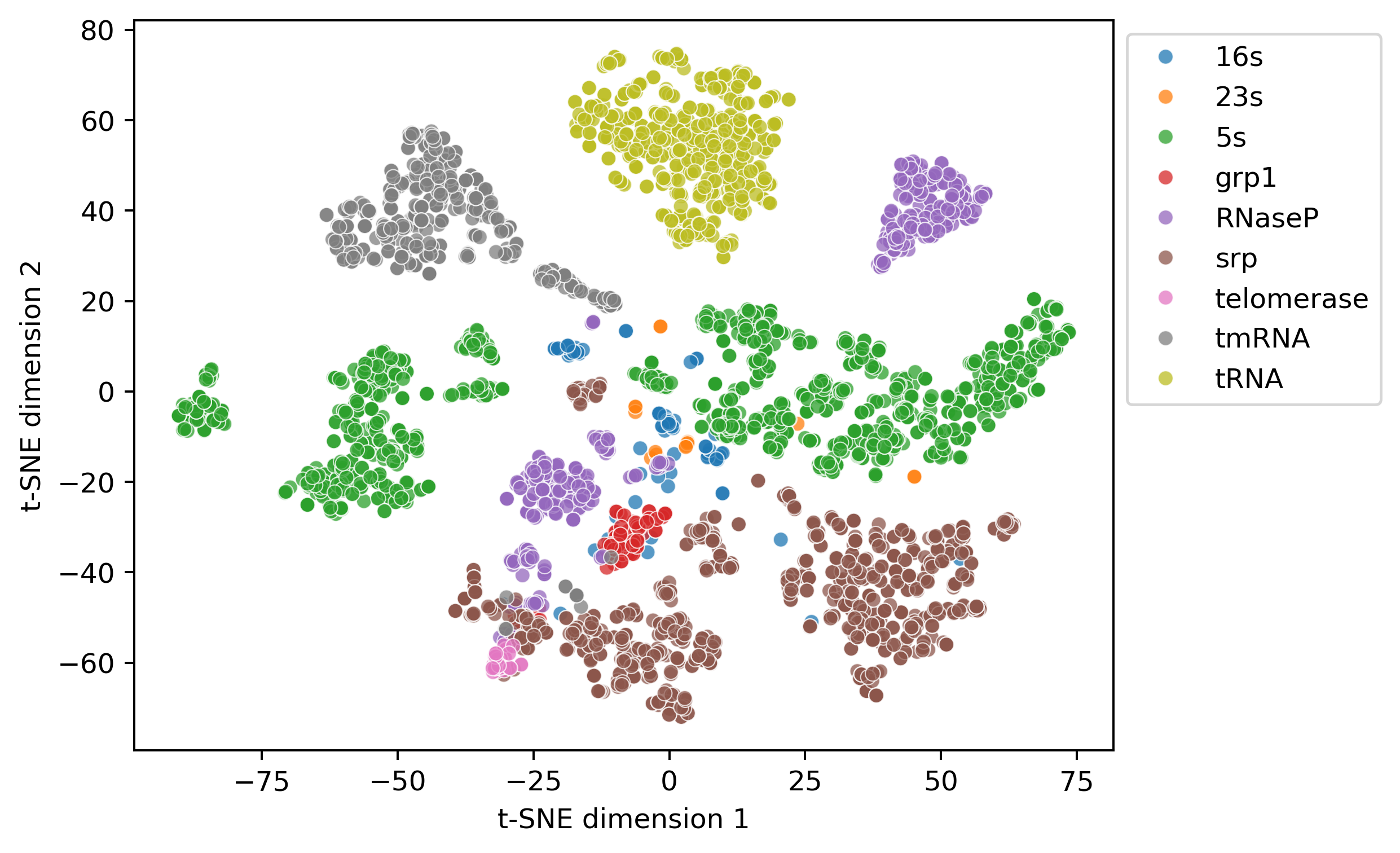}
    \caption{\textbf{t-SNE visualization of sequence embeddings outputted by RiNALMo for RNAs from the inter-family generalization evaluation dataset.}}
    \label{fig:ss_tsne_fams}
\end{figure*}

To visualize pre-trained RiNALMo's sequence representations, we applied {t-SNE} on the classification token embeddings for RNAs from a secondary structure prediction dataset (see Fig.~\ref{fig:ss_tsne_fams}).
In the embedding space, the RNAs are clustered by families with, in general, clean boundaries between clusters. This shows RiNALMo's ability to cluster and distinguish different RNA families and confirms it can be used in various clustering analyses of RNA sequences. The {t-SNE} visualization is important from the RNA structure perspective as well since the RNAs from the same families fold similarly, meaning the structure information is also contained in the embedding space.

\begin{figure*}[!ht]
    \centering
    \includegraphics[width=\textwidth]{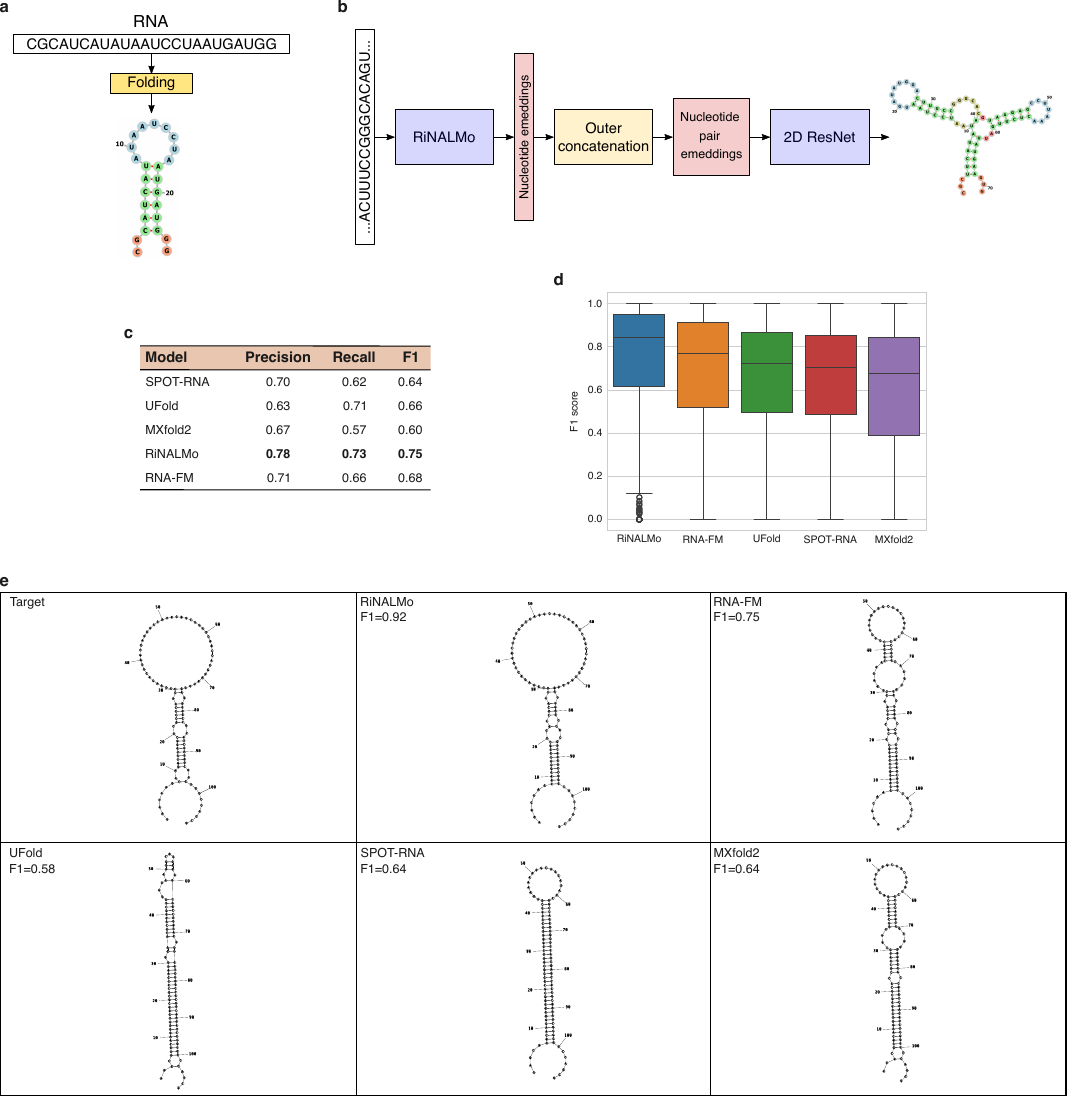}
    \caption{\textbf{Intra-family secondary structure prediction.} \textbf{a}, RNAs fold into various shapes according to their function and while doing so, many of their nucleotides pair up using a hydrogen bond. These pairings are crucial for structure stability and form structural motifs such as hairpin loops and bulges. \textbf{b}, RiNALMo produces nucleotide embeddings for the given RNA sequence. Nucleotide pair embeddings are constructed by applying outer concatenation to RiNALMo's outputs. Finally, pair representations are fed into the convolutional bottleneck residual neural network (ResNet) which produces base pairing probabilities that are then converted into the final secondary structure prediction. \textbf{c}, Precision, recall and F1 performance of different deep learning models on the TS0 evaluation dataset. \textbf{d}, Distribution of F1 scores for predictions of different models on the TS0 dataset. \textbf{e}, A target RNA from the TS0 evaluation dataset and its predictions from different deep learning models.}
    \label{fig:intra_fam_sec_struct_pred}
\end{figure*}

\bmhead{Fine-tuning RiNALMo for intra-family secondary structure prediction}
When RNAs fold into complex structures, many of their bases pair up and form hydrogen bonds. These pairs are vital for the structure's stability and function. These bonds can be represented by secondary structure, which can tell us a lot about RNA and which is often used as an input to the tertiary structure prediction tools. An example of a secondary structure can be seen in Fig. \ref{fig:intra_fam_sec_struct_pred}a. 

Most popular secondary structure prediction tools often rely on thermodynamic models, aiming to identify secondary structures that possess the lowest free energy \cite{reuter2010_rnastructure}. There are also popular probabilistic methods based on statistical learning procedures that act as an alternative to free energy minimization methods, such as CONTRAfold \cite{do2006_contrafold}. Several DL methods have been developed as well. They often outperform the thermodynamic models on RNA families on which they were trained, i.e., on in-distribution data.

We fine-tuned RiNALMo on a simple binary classification task with binary cross-entropy loss, where we tasked the model to classify each nucleotide pair as either paired or unpaired. The pipeline for determining secondary structures is illustrated in Fig. \ref{fig:intra_fam_sec_struct_pred}b. We utilized a dataset proposed in \cite{singh_spotrna} and compared our model to RNA-FM \cite{chen2022_rna_fm} and popular DL methods specialized for secondary structure prediction SPOT-RNA \cite{singh_spotrna}, UFold \cite{fu2021_ufold} and MXfold2 \cite{Sato2021_mxfold2}. Proposed training and evaluation datasets are denoted as TR0 and TS0, respectively, and are non-redundant to each other on the sequence similarity level. However, notice that sequences from the same RNA family can appear in both the training and evaluation datasets. All models were trained on the same training dataset (TR0) except SPOT-RNA which was additionally fine-tuned on a smaller dataset derived from PDB \cite{berman2000_pdb}. As can be seen in Fig. \ref{fig:intra_fam_sec_struct_pred}c, RiNALMo outperforms other state-of-the-art DL approaches in terms of precision, recall and consequently F1 score. We provide F1 score distributions in Fig. \ref{fig:intra_fam_sec_struct_pred}d. A TS0 target example and the predictions from different DL methods are given in Fig. \ref{fig:intra_fam_sec_struct_pred}e.

\bmhead{RNA language models can generalize well on inter-family structure prediction tasks}
While DL methods for secondary structure prediction outperform thermodynamic models on in-distribution data, they are usually unable to generalize well on new RNA families. This is a severe limitation as it hinders practical usage of such tools \cite{szikszai2022_deep_dont_generalize}. 

To test the generalization capabilities of RiNALMo, we utilized the dataset proposed in \cite{szikszai2022_deep_dont_generalize}. The dataset of $3865$ RNAs from nine families was split nine times, and in each split, a different family was held out for evaluation while the other eight families were used for training and validation. RiNALMo's ability to generalize across different RNA families was compared to RNA-FM \cite{chen2022_rna_fm}, popular thermodynamics-based tool RNAstructure \cite{reuter2010_rnastructure}, widely used probabilistic method CONTRAfold \cite{do2006_contrafold}, and two DL approaches specialized for secondary structure prediction UFold \cite{fu2021_ufold} and MXFold2 \cite{Sato2021_mxfold2}. The LMs were fine-tuned and the other two DL models were separately trained on each of the previously described dataset splits and evaluated on a corresponding unseen RNA family. For predicting the secondary structures using CONTRAfold, we used EternaFold parameters \cite{wayment2022_eternafold} trained on the EternaBench dataset, a set of more than $20,000$ RNAs.

\begin{figure*}[t]
    \centering
    \includegraphics[width=\textwidth]{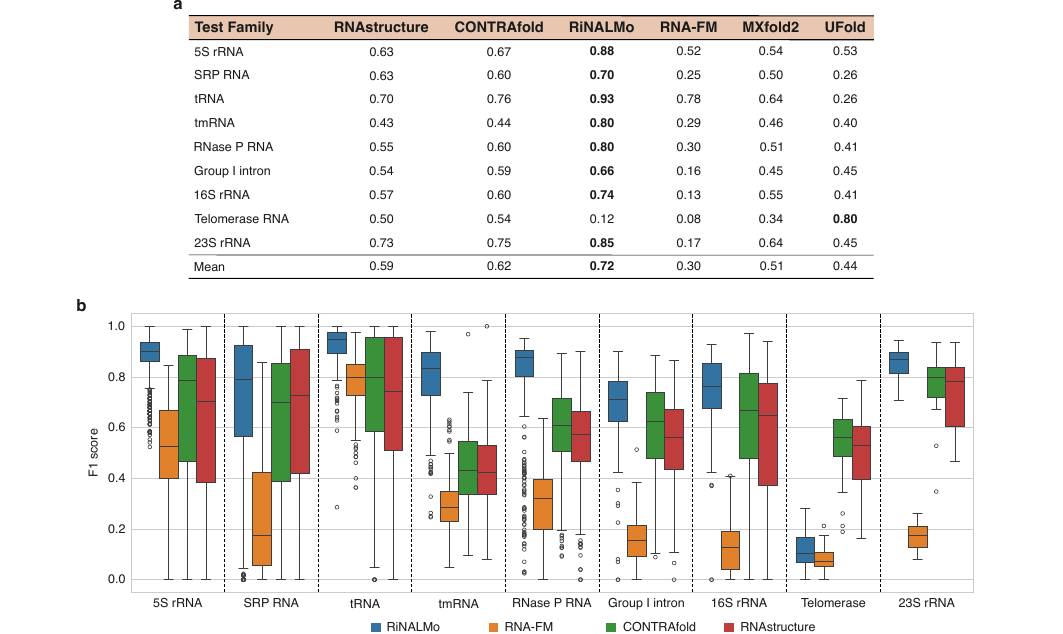}
    \caption{\textbf{Inter-family secondary structure prediction.} \textbf{a}, Secondary structure prediction average F1 scores for the ArchiveII evaluation datasets. \textbf{b}, Distribution of F1 scores for different methods on the ArchiveII evaluation datasets.}
    \label{fig:archive_sec_struct_pred}
\end{figure*}

Average F1 scores and F1 score distributions for different dataset splits are shown in Fig \ref{fig:archive_sec_struct_pred}.
Fine-tuned RiNALMo demonstrates that it is capable of inter-family generalization as it outperforms RNAstructure and CONTRAfold in eight out of nine families by high margins, unlike other DL models. To the best of our knowledge, this is the first paper to show that LMs can generalize well on inter-family secondary structure prediction, mitigating the limitations of other DL methods. We noted, however, that RiNALMo struggles to generalize on telomerase RNAs, but it achieves the highest F1 score on all other families. Visualization of RiNALMo's sequence embeddings for all RNAs in the dataset is presented in Fig. \ref{fig:ss_tsne_fams}. One can notice that telomerase RNAs are clustered together, however, there is no clear boundary between them and SRP RNAs. We also noticed that telomerase RNAs are the longest in the dataset, on average around $25\%$ longer than the second-longest in the dataset. Interestingly, UFold performs best on telomerase RNA, while achieving much worse results on the other families. We are currently unable to conclude why RiNALMo fails on telomerase RNAs, but will take more focus on this problem in the future. Technical details and more secondary structure prediction results and examples can be found in Methods, Supplementary materials and Extended Data Fig. \ref{fig:ss_examples_archiveII}.

\begin{figure*}[t]
    \centering
    \includegraphics[width=\textwidth]{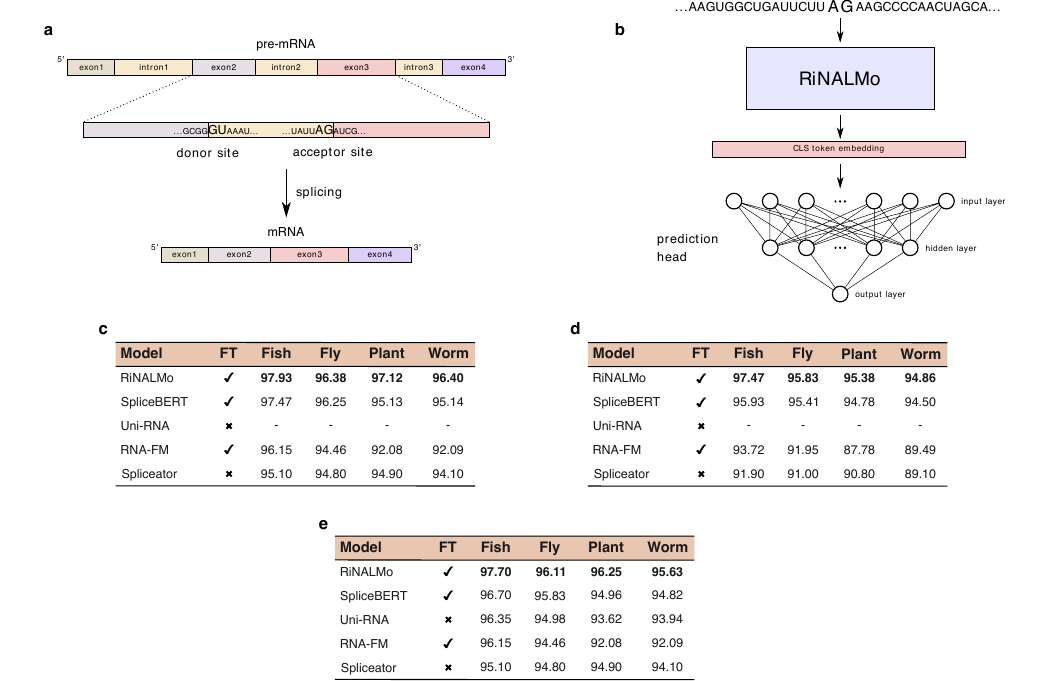}
    \caption{\textbf{RNA splice-site prediction.} \textbf{a}, A pre-mRNA transcript consists of non-coding, i.e., introns, and coding regions, i.e., exons. Introns are located between two exons of a gene. As part of the RNA processing pathway, introns are removed by cleavage at splice sites. These sites are found at $5'$ and $3'$ ends of introns, known as donor and acceptor splice sites, respectively. Most frequently, the $5'$ end of introns begins with the dinucleotide GU, and the $3'$ end of introns ends with AG. \textbf{b}, An input to RiNALMo is a $400$ nucleotide long RNA sequence from the GS\_1 dataset. RiNALMo outputs $402$ embeddings including the CLS embedding. We utilize only the CLS embedding that then passes through a two-layer MLP prediction head featuring a hidden layer with $128$ dimensions and employing the GELU activation function. The output layer gives information on whether a sequence contains a donor/acceptor site or not. \textbf{c}, Classification F1 score for donor splice-site prediction. \textbf{d}, Classification F1 score for acceptor splice-site prediction. \textbf{e}, Classification F1 score for splice-site prediction. Here, we report the average value of donor and acceptor prediction results. In \textbf{c}, \textbf{d} and \textbf{e}, FT denotes whether we fine-tuned the model or represented direct citations from original papers with the same split train/test datasets. ``-" denotes that the original paper did not report the result.}
    \label{fig:splice_site_pred}
\end{figure*}

\bmhead{Fine-tuning RiNALMo for splice-site prediction}
RNA splicing plays an important role in eukaryotic gene expression, involving the removal of introns from pre-mRNAs and the ligation of exons to form mature mRNAs (see Fig. \ref{fig:splice_site_pred}a). Precisely pinpointing splice sites—the donor and acceptor sites that mark the boundaries between exons and introns, and vice versa—is essential for accurately predicting gene structure and location.

Identifying the splice sites can be cast as a binary sequence-level classification task. A widely used dataset of positive and negative subsets of splice-site sequences was proposed in \cite{scalzitti2021_spliceator}. The dataset was constructed by randomly selecting sequences from the exon/intron regions of the G3PO+ genomic sequences \cite{scalzitti2020benchmark}. The dataset consists of error-free splice-site sequences from a diverse set of $148$ eukaryotic organisms, including humans. The test dataset consists of four different species not seen in the training dataset.

We separately fine-tuned the model, first for donor and then for acceptor splice-site prediction. The splice-site prediction pipeline using RiNALMo embeddings is illustrated in Fig. \ref{fig:splice_site_pred}b. Finally, we compared our model's performance with other RNA LMs RNA-FM \cite{chen2022_rna_fm} and Uni-RNA \cite{Wang2023_unirna}, and several established methods such as Spliceator \cite{scalzitti2021_spliceator} and SpliceBERT \cite{Chen2023_splice_bert}. We present the results in Fig. \ref{fig:splice_site_pred}.
We fine-tuned other models and used the same prediction head if they were publicly available and easy to fine-tune. Our fine-tuned model outperforms other models, showing its powerful generalization properties. Notice that RiNALMo even outperforms SpliceBERT, an LLM pre-trained exclusively on pre-mRNA sequences. More details on the splice-site prediction task and the model's hyperparameters can be found in Methods and Supplementary materials.

\bmhead{Mean ribosome loading prediction using fine-tuned RiNALMo}
Due to efficiency reasons, cells usually use groups of multiple ribosomes to translate the mRNA. These groups are called polyribosomes, enabling the cell to create multiple proteins from a single mRNA. To quantify protein synthesis activity, an MRL metric, defined as the average number of ribosomes that translate the mRNA instructions into polypeptides, has been introduced.

MRL prediction task can be viewed as a regression task where the input is the $5'$ UTR region of the mRNA. Commonly used datasets of $5'$ UTR sequences with measured MRL values are provided by \citet{Sample2019}. Two evaluation datasets, namely Random7600 and Human7600, were created by sampling the original dataset containing human and random UTR sequences.

\begin{figure*}[t]
    \centering
    \includegraphics[width=\textwidth]{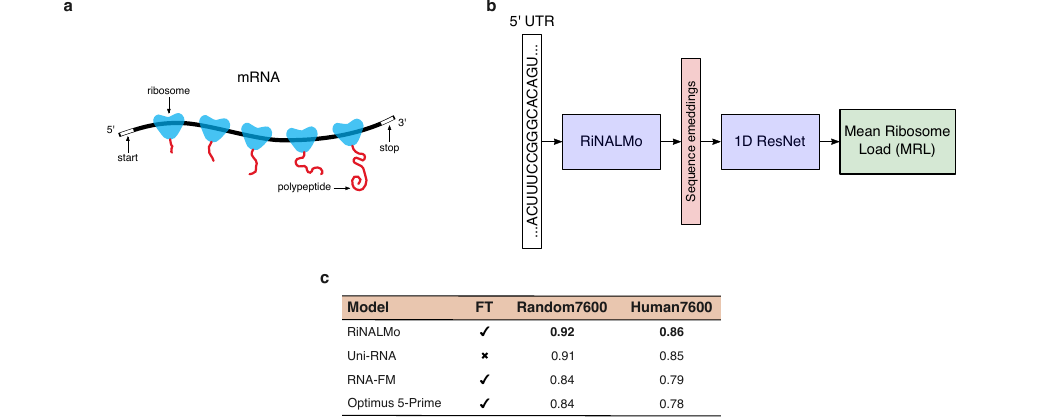}
    \caption{\textbf{Mean ribosome loading prediction.} \textbf{a}, To produce multiple proteins from a single mRNA strand, cells utilize multiple ribosomes. The mean ribosome load (MRL) metric was designed to indicate the translation efficiency of a certain mRNA. \textbf{b}, Output embeddings produced by RiNALMo are forwarded to the one-dimensional convolutional ResNet which outputs the mean ribosome load prediction. \textbf{c}, $R^2$ score for MRL prediction on the Random7600 and Human7600 datasets. FT denotes whether we trained the model or represented direct citations from the original papers with the same split train/test datasets.}
    \label{fig:mrl_pred}
\end{figure*}

We fine-tuned RiNALMo to predict the MRL value for the given $5'$ UTR sequence. The prediction pipeline is illustrated in Fig. \ref{fig:mrl_pred}b. RiNALMo outputs are fed into a prediction head consisting of six ResNet blocks. The mean squared error was used as the loss function. The model's performance was compared to other RNA LMs Uni-RNA \cite{Wang2023_unirna} and RNA-FM \cite{chen2022_rna_fm}. We also compared our model to the popular Optimus 5-prime model \cite{Sample2019} specialized for MRL prediction. $R^2$ was used as the evaluation metric, and the results are reported in Fig. \ref{fig:mrl_pred}c.
Fine-tuned RiNALMo outperforms other models. Notice that RiNALMo can generalize on human UTRs despite being fine-tuned only on sequences of random origin, again proving its generalization capability. More details on MRL prediction can be found in Methods and Supplementary materials.

\section{Discussion}
\label{sec:discussion}
We pre-trained our RNA language model on a vast amount of ncRNA sequences from several databases, including RNAcentral, Rfam, nt, and Ensembl. The data 
was carefully curated to ensure sequence diversity in each training batch, which subsequently led to better generalization capabilities of the pre-trained language model. We used t-SNE on the classification token embeddings to show that RiNALMo's output representations contain information about the RNA family and that the RNAs with similar structures are close in the model's embedding space. A more insightful way of assessing the embeddings' expressiveness and whether the model captures hidden structure information is to assess it on downstream tasks.

First, we fine-tuned and tested RiNALMo on two secondary structure prediction tasks. The first task was an intra-family secondary structure prediction, where RNAs from the same family come in both training and test datasets. The results showed that fine-tuned RiNALMo's output representation embeds important structure information about the input sequence and, when utilized with a proper prediction head, leads to state-of-the-art performance. The second task was an inter-family secondary structure prediction, where an RNA family from the test dataset was not seen in the training dataset. It was shown previously that deep learning methods do not generalize well across RNA families \cite{szikszai2022_deep_dont_generalize}. However, RiNALMo outperformed both thermodynamics-based and deep learning methods and showed that RNA language models can generalize well on unseen RNA families. This demonstrates the outstanding generalization capability of RiNALMo for secondary structure prediction tasks.

Furthermore, we fine-tuned RiNALMo on two function-related RNA downstream tasks. Both of these tasks were related to the mRNA family whose examples were not contained in the pre-training dataset. The results from Section~\ref{sec:results} showed that RiNALMo again generalizes well and can capture important functional information from RNA sequences from previously unseen RNA types.

In future work, we will explore the augmentation of the pre-training dataset by adding coding RNAs and evaluate how it affects the model performance on the structural and functional downstream tasks. Having bigger data might require a larger RNA LM which we also have in consideration. Multimodal pre-training including sequence and chemical mapping data is another direction for improving the representations for structure-related downstream tasks which we will explore in the future. Finally, we plan to employ RiNALMo in several other structure and function-related tasks. Of particular interest would be to test RiNALMo on tertiary structure prediction tasks and see whether its expressive output embeddings and generalization capability can improve the performance of existing or new prediction tools. Another interesting application would be employing RiNALMo's embeddings for sequence conditioning for RNA design \cite{anand2024rna}.

To conclude, we presented RiNALMo, a new largest-to-date general-purpose RNA LM pre-trained on a dataset of $36$M ncRNA sequences using MLM. We showed that by pre-training on a carefully curated RNA dataset and using the most modern Transformer techniques, an LM can capture hidden knowledge and important structural information from unlabeled RNA sequences. The results on downstream tasks proved the generalization capability of RiNALMo and the expressiveness of its output representations. Of particular importance are the results of the inter-family secondary structure prediction task, where we showed that RiNALMo can generalize well on RNA families unseen during fine-tuning, unlike other DL methods. Due to the significance of the results, we believe RiNALMo presents a valuable asset for advancing our understanding of RNA structures and functions.

\clearpage

\section{Methods}\label{sec11}
\subsection{RNA Language Model}
RiNALMo is an encoder-only Transformer. An input RNA sequence is tokenized, turned into a $1280$ dimension vector using a learned input embedding model and passed to the Transformer. RiNALMo consists of $33$ Transformer blocks, and each block comprises a multi-head attention and a feed-forward network (FFN). The position of the tokens is encoded using the RoPE \cite{su2024_roformer}, which effectively encapsulates both the relative and the absolute positional information. Each multi-head attention has $20$ attention heads. The multi-head attention is illustrated in Supplementary materials. To improve the pre-training efficiency of such a large model, we employed the IO-aware FlashAttention-2 \cite{dao2023_flashattention2}, a fast and memory-efficient exact attention. In the FFN, we use two linear layers and the SwiGLU activation function \cite{shazeer2020_swiglu} which combines the advantages of the Swish activation function and the gated linear unit (GLU). The FFN layers have hidden size $d_{ff}=3413$, scaling the model to $650$M parameters. The Transformer modules are interconnected using residual connections. We use the layer normalization put inside the residual blocks to stabilize the training and have well-behaved gradients at initialization \cite{xiong2020layer}. We did an ablation study for RoPE and the SwiGLU activation function which showed the effectiveness of these techniques on the downstream tasks compared to the conventional absolute positional encoding and GELU activation function used in RNA-FM. The ablation study results are provided in Supplementary materials.

\subsection{Tokenization}
In this work, each nucleotide is a single token. During tokenization, we replace all ``U"s in the sequences with ``T"s. This leads to a vocabulary involving four main types of nucleotides (``A", ``C", ``T" and ``G") and several other types of combinations (``I", ``R", ``Y", ``K", ``M", ``S", ``W", ``B", ``D", ``H", ``V", ``N", and ``-"). We use additional tokens used commonly in MLM, such as \verb|[CLS]|, \verb|[EOS]|, \verb|[PAD]|, and \verb|[MASK]|. During masking, we change nucleotides with the \verb|[MASK]| token or replace it with one of the four main types of nucleotides from the vocabulary or the ``any nucleotide" (``N") token. Tokens \verb|[CLS]| and \verb|[EOS]| are added at the beginning and end of the sequence. The \verb|[PAD]| token is appended at the end of shorter sequences to have all the sequences in a batch of the same length.

\subsection{Data Preprocessing}
For pre-training dataset preprocessing, we implemented a multi-step preparation pipeline. First, we collected non-coding RNA sequences from publicly available datasets RNAcentral, nt, Rfam and Ensembl. We removed sequences shorter than $16$ and longer than $8192$. Furthermore, we also removed sequence duplicates with \textit{seqkit rmdup} and the resulting unique sequences were clustered with \textit{mmseqs easy-linclust} with options \textit{-{}-min-seq-id $0.7$} and \textit{-c $0.8$}. Finally, we saved the processed dataset into \textit{LMDB} (Lightning Memory Mapped Database) so we could easily and quickly access any data sample during the pre-training. In the end, the dataset consisted of $36$M unique ncRNA sequences clustered into $17$M clusters. In comparison, \citet{chen2022_rna_fm} collected ncRNAs from the RNAcentral database only. As a preprocessing step, the authors removed only duplicate RNA sequences. We clustered the RNA sequences to ensure sequence diversity in each training batch since we later sample from each cluster once per epoch.

\subsection{Pre-training}
We pre-trained RiNALMo using the MLM task where we corrupted unlabeled RNA sequences and then tasked the model to reconstruct them. To corrupt the input sequence, we randomly selected $15\%$ of the tokens in the training sequence. Of these, $80\%$ are masked, i.e., replaced with the unique vocabulary token \verb|[MASK]|, $10\%$ are replaced with a randomly selected token from the vocabulary, and the remaining $10\%$ are left intact.

Let us denote the corrupted sequence by ${\tilde{\bm{X}}=\{\tilde{x}_i\}}$. We train RiNALMo parameterized by $\theta$ to reconstruct ${\bm{X}}$ by predicting the masked tokens conditioned on $\tilde{\bm{X}}$. For a given input token $\tilde{x}_i$, the loss is the probability of the correct nucleotide, given $\tilde{\bm{X}}$:
\begin{equation*}
    \mathcal{L}_{MLM}= -\log p_{\theta}(\tilde{x}_i = x_i | \tilde{\bm{X}}).
\end{equation*}
The weight updates are based on the average loss over the sampled tokens from a single training sequence (or a batch of sequences):
\begin{equation*}
    \mathcal{L}_{MLM}= -\frac{1}{|\mathcal{M}|} \sum_{i \in \mathcal{M}} \log p_{\theta}(\tilde{x}_i = x_i | \tilde{\bm{X}}),
\end{equation*}
where $\mathcal{M}$ is the index set of masked tokens.

Due to the computational limitations, during training, we limited the length of the input into the language model to $1024$ tokens. Every sequence begins with a \verb|[CLS]| token, i.e., a classification token, and ends with an \verb|[EOS]| token, i.e., an end-of-sequence token. If an input sequence is longer than $1022$ nucleotides, a random part of the sequence will be cropped out and fed into the model during each epoch, similar to what was done in \cite{rives2021_esm_1b}. To implicitly give information to the model that the sequence was cropped, it does not begin with the \verb|[CLS]| token nor ends with the \verb|[EOS]| token unless we crop the beginning or end of the sequence.

To ensure sequence diversity in each training batch during pre-training, we randomly sampled each sequence in the batch from a different sequence cluster. Effectively, it means that in every epoch RiNALMo saw $17$M samples, i.e., one sequence from each of the clusters, and in each epoch, sequences were sampled randomly from the clusters using a new seed.

We pre-trained RiNALMo with seven A100 GPUs of $80$ GB memory for two weeks. The batch size was set to $192$ per GPU. We adopted the cosine annealing learning rate schedule with a linear warm-up. During the warm-up period, learning rate increases from $10^{-7}$ to $5 {\times} 10^{-5}$ for $2000$ steps. For the cosine annealing schedule, the minimum learning rate was set to $10^{-5}$. The gradient norm was clipped to 1.0.

We pre-trained several configurations of the LM: RiNALMo-$650$M with $650$M parameters, RiNALMo-$150$M with $148$M parameters, and RiNALMo-$33$M with $33.5$M parameters. We did several experiments to evaluate how the model's size influences its performance on the pre-training and downstream tasks. First, we measured perplexity. Perplexity is a metric used to evaluate how well an LM predicts a sample of text or, in our case, an RNA sequence. The perplexity is defined as the exponential of the negative log-likelihood of the sequence. To efficiently calculate the perplexity we used the following term
\begin{equation*}
    \textsc{Perplexity}(x) = \exp\{ -\log p_\theta(\tilde{x}_{i \in \mathcal{M}} = x_{i} | \tilde{\bm{X}})\}.
\end{equation*}
Lower perplexity means the LM is better at reconstructing the masked tokens in the MLM pre-training task.

In order to evaluate how the number of parameters affects the model's ability to reconstruct the masked tokens, we calculated the perplexity of RiNALMo for several different model sizes. Validation perplexity was measured on a 1\% random-split holdout of the pre-training dataset. The perplexity values are given in Extended Data Fig.~\ref{fig:perplexity}. It can be seen that as we increase the model size, the perplexity decreases, meaning the larger the LM gets, it is more capable of reconstructing the missing tokens.

The hyperparameters of RiNALMo models of different sizes and the comparison of their performance on the downstream tasks are given in Supplementary materials. Based on these results we can conclude that besides novel techniques such as RoPE and SwiGLU, model size is important in achieving a boost in performance on pre-training and consequently downstream tasks.

\subsection{Secondary Structure Prediction}
To adapt RiNALMo for the secondary structure prediction, we fine-tuned the model with a simple binary classification task with binary cross-entropy loss, where we tasked the model to classify each nucleotide pair as either paired or unpaired. A simple bottleneck residual neural network (ResNet) \cite{he2016_resnet} is attached to the language model and is fed with nucleotide pair representations. These representations are obtained by applying outer concatenation to RiNALMo's output. For example, to obtain the vector representation of the nucleotide pair $(i, j)$, we simply concatenate the representation of the nucleotide $j$ to the representation of the nucleotide $i$.

In the end, our secondary structure prediction pipeline outputs a matrix where each element represents a pairing probability logit for a certain nucleotide pair. Because of the symmetry of secondary structures (if nucleotide $i$ is paired with nucleotide $j$, then $j$ is paired with $i$ as well), we calculate training loss only on matrix elements ``above" the main diagonal.

During fine-tuning, we utilized gradual parameter unfreezing. After every three epochs, we unfroze additional three RiNALMo's layers. In the first three epochs, we trained only the prediction head. The model was fine-tuned for $15$ epochs and the learning rate was initially set to $10^{-4}$ with a linear decay schedule. We used the same prediction head architecture for RNA-FM fine-tuning. Due to architectural differences between RiNALMo and RNA-FM, we decided to modify the fine-tuning schedule. The prediction head was first pre-trained while RNA-FM parameters were kept frozen. After three epochs we unfroze RNA-FM and fine-tuned it alongside the prediction head.

To convert base pairing probabilities to a proper secondary structure, we implemented a simple greedy approach where we iteratively set nucleotide pairs with the highest pairing probability as paired and then excluded all possible clashing pairs from being set as paired in future iterations of the algorithm. During this procedure, we ignore non-canonical nucleotide pairings and pairings that would cause a "sharp" hairpin loop ($i$-th nucleotide cannot be paired with the $j$-th nucleotide if $|i - j| < 4$). The classification threshold was tuned on the validation set to ensure a balanced pairing ratio.

When assessing the performance of secondary structure prediction, it is important to consider RNA structural dynamics, and that's why it is helpful to view predictions that are close enough as correct. We employed the metric calculation approach proposed in \cite{mathews2019_eval_sec_struct}. To be more precise, for a nucleotide pairing $(i, j)$ where $i$ and $j$ represent nucleotide indices in the RNA sequence, $(i \pm 1, j)$ and $(i, j \pm 1)$ pairings are also considered correct predictions. To obtain the F1 scores we reported in this study, we calculated the F1 score separately on each structure and then averaged those values.

In Supplementary materials, we provide the performance of smaller LMs, the ablation study and the impact of fine-tuning. We also report sequence-length weighted F1 scores on the inter-family secondary structure prediction task.

\subsection{Multi-Species Splice-Site Prediction}
Splice-site prediction task is essentially a binary sequence-level classification task, where a method has to detect whether a query RNA sequence contains a donor/acceptor site or not. We fine-tuned RiNALMo on the training dataset to predict splice sites in the sequences. The classification token (CLS) was used as input to the prediction head. This head is configured as a two-layer multilayer perceptron (MLP) featuring a hidden layer with 128 dimensions and employing the GELU activation function. Cross-entropy loss served as our choice for the loss function. We separately fine-tuned the model, first for the donor and then for the acceptor splice-site prediction task. 

We used the dataset denoted as GS\_1 in the paper it was proposed in \citet{scalzitti2021_spliceator}. GS\_1 has the same ratio of positive to negative samples, whose negative samples consist of exon, intron, or false positive sequences. The length of the input sequences was set to $400$ nucleotides for both the training and test datasets. 

The prediction head was uniformly initialized from $\mathcal{U}(-\sqrt{1/d}, \sqrt{1/d})$, where $d$ is the input features dimension. The learning rate was set to $10^{-5}$ and the language model with the prediction head was fine-tuned for $2$ epochs when it achieved the best results on the validation dataset. The batch size was set to $32$.

We fine-tuned the SpliceBERT model and the RNA-FM model in combination with a two-layer MLP prediction head. We fine-tuned the models on the same training/validation dataset used for fine-tuning RiNALMo with the same learning rate.

More splice-site prediction results can be found in Supplementary materials.

\subsection{Mean Ribosome Loading Prediction}
We fine-tuned RiNALMo on the appropriate training dataset to predict the MRL value for the given $5'$ UTR sequence. Language model outputs are fed into a prediction head which consists of six ResNet convolutional blocks. The mean squared error was used as the loss function. RNA-FM was fine-tuned with the same prediction head as RiNALMo.

Each block of the prediction head consists of two 1D convolution layers followed by instance normalization and ELU activation function. Parameters of the prediction head were initialized with the default Pytorch \cite{paszke_pytorch} parameter initialization method.

Training and evaluation datasets were produced with the procedure described in \cite{Sample2019}. Two evaluation datasets were created by sampling the original dataset that contains human and random UTR sequences of varying lengths. To ensure that each sequence length is represented equally in the dataset, $100$ sequences with the deepest read coverage were selected for every length from $25$ to $100$ nucleotides. The same approach has been applied to both the human and random $5'$ UTR sequences, yielding two evaluation datasets called Random7600 and Human7600. All remaining random $5'$ UTRs with acceptable read coverage depth were used as the training dataset.

All MRL targets were standardized with the mean and standard deviation of training MRL values.

The model was fine-tuned for $50$ epochs. In the first $5$ epochs of the training, only the prediction head was trained. The learning rate was set to $10^{-4}$ and linearly decayed to $10^{-5}$ over the first 5000 training steps after which it remained constant. The batch size was set to 64. RNA-FM was fine-tuned with the same training procedure.

More MRL prediction results can be found in Supplementary materials.

\section{Data availability}
\label{sec:data_avail}
We used several databases with unannotated RNA sequences, namely RNAcentral  \cite{rnacentral2021}, nt \cite{nt_2023}, Rfam \cite{kalvari_rfam2020} and Ensembl \cite{ensembl_2023}. The RNAcentral dataset is available at \href{https://ftp.ebi.ac.uk/pub/databases/RNAcentral/current_release/sequences/}{https://ftp.ebi.ac.uk/pub/databases/RNAcentral/current\_release/sequences/}. The nt dataset is available at \href{https://ftp.ncbi.nlm.nih.gov/blast/db/FASTA/}{https://ftp.ncbi.nlm.nih.gov/blast/db/FASTA/}. The Rfam dataset is available at \href{https://ftp.ebi.ac.uk/pub/databases/Rfam/CURRENT/fasta_files/Rfam.fa}{https://ftp.ebi.ac.uk/pub/databases/Rfam/CURRENT/fasta-files/Rfam.fa}. Finally, the Ensembl dataset is available at \href{https://ftp.ensembl.org/pub/release-109/fasta/}{https://ftp.ensembl.org/pub/release-109/fasta/}.

The intra-family RNA secondary structure dataset with train/test splits is available at \href{https://dl.dropboxusercontent.com/s/w3kc4iro8ztbf3m/bpRNA_dataset.zip}{https://dl.dropboxusercontent.com/s/w3kc4iro8ztbf3m/bpRNA\_dataset.zip}. The inter-family secondary structure dataset consisting nine families and train/test splits is available at \href{https://github.com/marcellszi/dl-rna/releases/download/Data/ct-splits.tar.gz}{https://github.com/marcellszi/dl-rna/releases/download/Data/ct-splits.tar.gz}.

For the multi-species splice-site downstream task, we used the dataset denoted as GS\_1 in \cite{scalzitti2021_spliceator}. The dataset is available at \href{https://git.unistra.fr/nscalzitti/spliceator}{https://git.unistra.fr/nscalzitti/spliceator} and \href{https://zenodo.org/records/7995778}{https://zenodo.org/records/7995778}.

The mean ribosome loading dataset \cite{Sample2019} used in the paper can be found at \href{https://www.ncbi.nlm.nih.gov/geo/download/?acc=GSE114002&format=file}{https://www.ncbi.nlm.nih.gov/geo/download/?acc=GSE114002\&format=file}.

\section{Code availability}
The code repository is available on \href{https://github.com/lbcb-sci/RiNALMo}{https://github.com/lbcb-sci/RiNALMo} and the pre-trained and fine-tuned weights are available on \href{https://zenodo.org/records/10725749}{https://zenodo.org/records/10725749}. The weights can be automatically downloaded using the script provided in the repository. We provide scripts for fine-tuning the pre-trained model on the downstream tasks from the Results section. The data used in the downstream tasks can be automatically downloaded and preprocessed using the scripts in the code repository.

\backmatter
\bibliography{main}

\section*{Acknowledgements}
The authors would like to thank Ivona Martinović for the valuable comments and fruitful discussion on this work.

This work was supported in part by the National Research Foundation (NRF) Competitive Research Programme (CRP) under Project \textit{Identifying Functional RNA Tertiary Structures in Dengue Virus} (NRF-CRP27-2021RS-0001) and in part by the A*STAR under Grant \textit{GAP2: A*STAR RNA-Foundation Model (A*STAR RNA-FM)} (I23D1AG079).

The computational work for this article was partially performed on resources of the National Supercomputing Centre, Singapore (\url{https://www.nscc.sg}).

\section*{Author contributions}
R.J.P. and M.Š. conceived the project. R.J.P. and T.V. designed the method. R.J.P. and T.V. designed and conducted the numerical experiments. R.G.H., Y.W. and M.Š. supervised the study. R.J.P. and T.V. wrote the manuscript. R.G.H, Y.W. and M.Š. provided mentorship and support during the project. All authors approved the manuscript.

\section*{Competing interests}
The authors declare no competing interests.

\section*{Additional information}
\bmhead{Extended data} Extended data are available in the appendix.

\bmhead{Supplementary information} The paper contains supplementary material.

\begin{appendix}
\section*{Appendix}
\label{app}

\makeatletter
\renewcommand{\figurename}{Extended Data Fig.}
\makeatother
\setcounter{figure}{0}

\begin{figure*}[!ht]
    \centering
    \includegraphics[width=\textwidth]{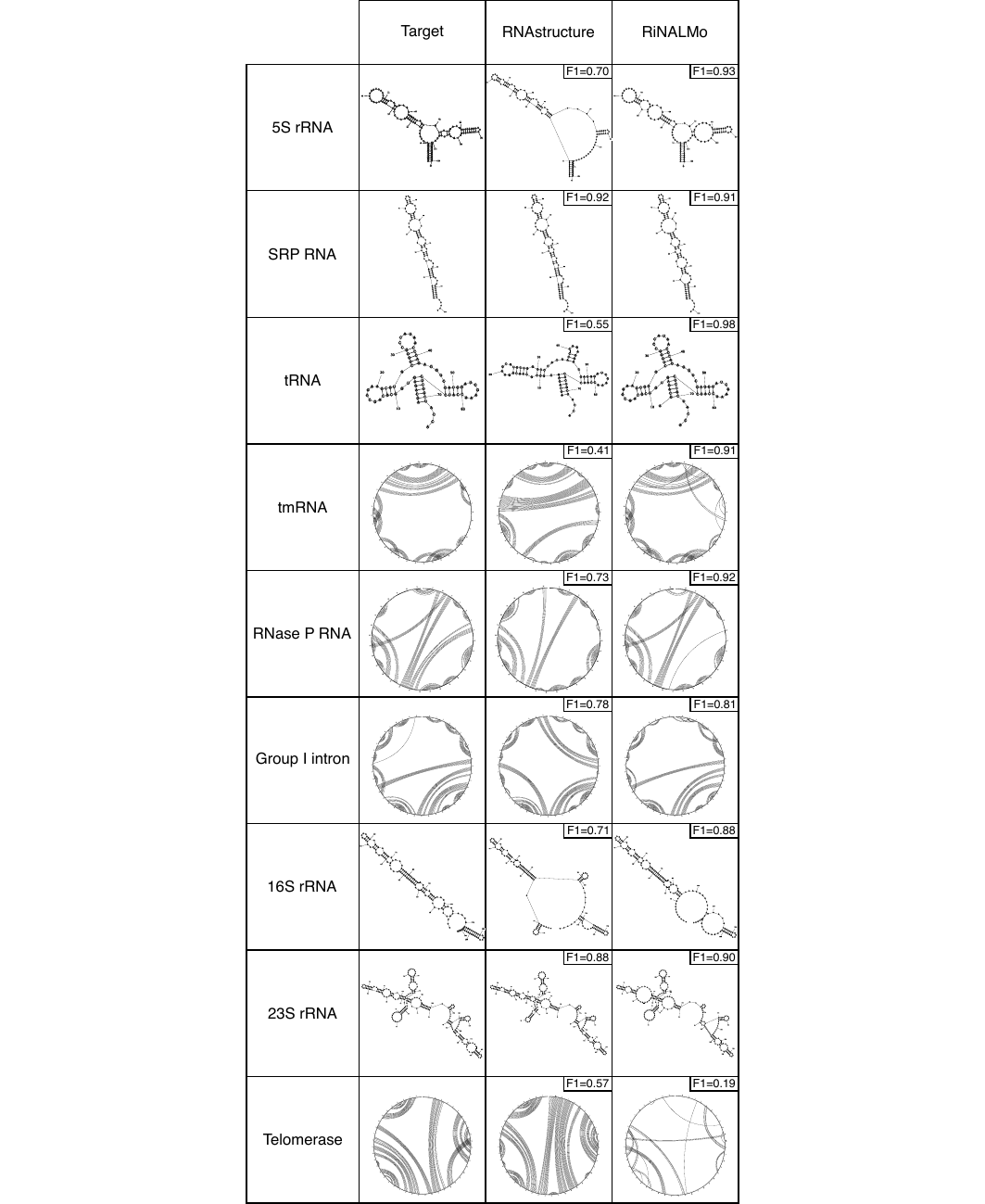}
    \caption{\textbf{Secondary structure prediction examples for the inter-family dataset.} Target secondary structures and predictions by RNAstructure and RiNALMo. The F1 score is shown in the upper right corner of each predicted structure figure.}
    \label{fig:ss_examples_archiveII}
\end{figure*}

\begin{figure}[t]
    \centering
    \includegraphics[width=0.8\textwidth]{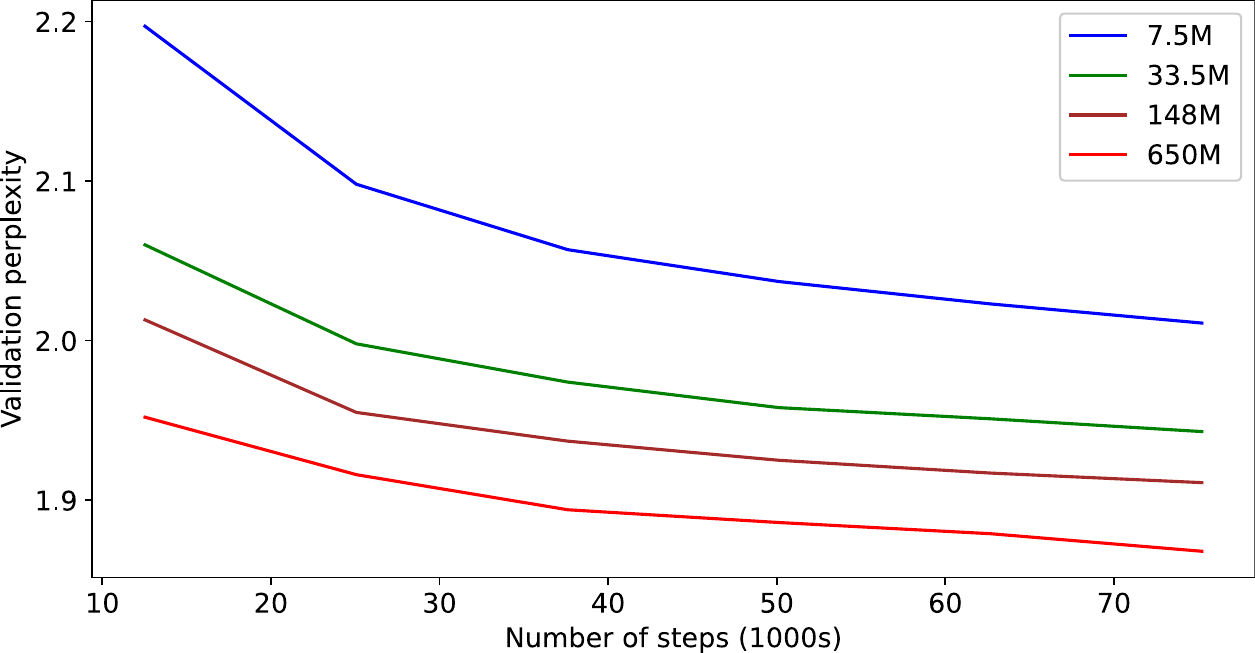}
    \caption{\textbf{RiNALMo masked language modeling perplexity curves.} Validation perplexity curves for RiNALMo models from $7.5$M to $650$M parameters. The models were trained for six epochs (around $77,000$ steps).}
    \label{fig:perplexity}
\end{figure}

\end{appendix}

\end{document}


\title[Supplementary Materials for RiNALMo]{Supplementary Materials for RiNALMo: General-Purpose RNA Language Models Can Generalize Well on Structure Prediction Tasks}

\author*[1]{\fnm{Rafael Josip} \sur{Penić}}\email{rafael-josip.penic@fer.hr}
\author*[2]{\fnm{Tin} \sur{Vlašić}}\email{tin\_vlasic@gis.a-star.edu.sg}
\author[3]{\fnm{Roland G.} \sur{Huber}}\email{rghuber@bii.a-star.edu.sg}
\author[2]{\fnm{Yue} \sur{Wan}}\email{wany@gis.a-star.edu.sg}
\author*[2]{\fnm{Mile} \sur{Šikić}}\email{mile\_sikic@gis.a-star.edu.sg}

\affil[1]{\orgdiv{Faculty of Electrical Engineering and Computing}, \orgname{University of Zagreb}, \orgaddress{\street{3 Unska Street}, \city{Zagreb}, \postcode{10000}, \country{Croatia}}}

\affil[2]{\orgdiv{Genome Institute of Singapore (GIS)}, \orgname{Agency for Science, Technology and Research (A*STAR)}, \orgaddress{\street{60 Biopolis Street, Genome}, \city{Singapore}, \postcode{138672}, \country{Republic of Singapore}}}

\affil[3]{\orgdiv{Bioinformatics Institute (BII)}, \orgname{Agency for Science, Technology and Research (A*STAR)}, \orgaddress{\street{30 Biopolis Street, Matrix}, \city{Singapore}, \postcode{138671}, \country{Republic of Singapore}}}

\maketitle

\makeatletter
\renewcommand\thesection{S\@arabic\c@section}
\renewcommand\thetable{S\@arabic\c@table}
\renewcommand \thefigure{S\@arabic\c@figure}
\makeatother

\section{RNA Language Model}
\label{supp:rna_lm}
\begin{figure}[t]
    \centering
    \includegraphics[width=0.65\textwidth]{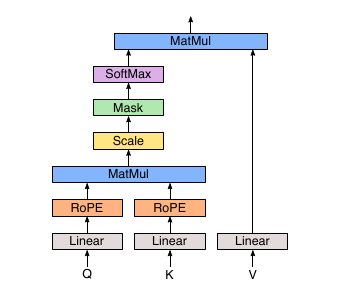}
    \caption{\textbf{RiNALMo's self-attention mechanism.} Instead of the traditional absolute positional encoding we employ rotary positional embedding (RoPE) that offers a more robust and scalable way of encoding positional information.}
    \label{fig:self_attention}
\end{figure}

We pre-trained several configurations of the language model (LM): RiNALMo-$650$M with $650$M parameters, RiNALMo-$150$M with $148$M parameters,  RiNALMo-$33$M with $33.5$M parameters. The configurations can be seen in Table \ref{tab:hyperparams}. The model is based on the self-attention mechanism. The self-attention diagram is illustrated in Fig. \ref{fig:self_attention}. The self-attention mechanism allows the model to identify and weigh the importance of different parts of the input RNA sequence by attending to itself. Model, pre-training and fine-tuning pipelines were implemented from scratch using \textit{Pytorch} and \textit{Pytorch Lightning} frameworks.

\begin{table}[t]
    \caption{Language model configurations and their hyperparameters.}
    \label{tab:hyperparams}
    \vskip 0.15in
    \begin{center}
        \begin{small}
            \begin{sc}
                \begin{tabular}{lccc}
                    \toprule
                    Hyperparameter & RiNALMo-33M & RiNALMo-150M & RiNALMo-650M  \\
                    \midrule
                    Layers   & $12$ & $30$  & $33$ \\
                    Embed. dim. & $480$ & $640$ & $1280$ \\
                    Attention heads & $20$ & $20$ & $20$ \\
                    RoPE & True & True & True \\
                    Attention dropout & $0.1$ & $0.1$ & $0.1$ \\
                    Residual dropout & $0.1$ & $0.1$ & $0.1$ \\
                    \bottomrule
                \end{tabular}
            \end{sc}
        \end{small}
    \end{center}
    \vskip -0.1in
\end{table}

\section{Secondary Structure Prediction}
\label{supp:sec_struct}
To assess the model's ability to accurately predict secondary structures for longer RNA sequences we also considered a weighted F1 score. To clarify, when calculating the average F1 score we weighted each score with the length of the corresponding RNA sequence. A comparison of weighted and non-weighted F1 scores for each RNA family during inter-family generalization evaluation is shown in Table \ref{tab:ss_f1_weight_ablation}. Notice that RiNALMo's performance drops only on longer SRP RNAs but holds up on longer sequences of all other families. The results show that on average RiNALMo's performance does not drop on longer sequences which is known to be a problem for several secondary structure prediction methods.

We compared RiNALMo with other state-of-the-art secondary structure prediction methods and the RNA-FM language model, however, we are missing a comparison with another language model Uni-RNA \cite{Wang2023_unirna}. Uni-RNA is not publicly available and we could not fine-tune it on the same training dataset we used in this paper. The only way to compare our model's performance to UNI-RNA is to replicate its training and test datasets. UNI-RNA study reports the usage of the bpRNA dataset (most probably SPOT-RNA's version) but also an expansion of the training set with the RNAStralign dataset. It is not clear if and how the added data was augmented in any way to prevent potential data leaks. If they did not do any filtering to their datasets, $79$ ($6\%$) RNA sequences from the test set are present in the training set which is a significant data leak. Because of the described uncertainties, we decided to leave out the comparison with Uni-RNA for the secondary structure prediction task.

In the paper, secondary structures have been visualized with \textit{forna}\footnote{\url{http://rna.tbi.univie.ac.at/forna/}} and RNAstructure's \textit{draw}\footnote{\url{https://rna.urmc.rochester.edu/Text/draw.html}} utility tool.

\begin{table*}[t]
    \caption{Inter-family generalization evaluation sequence length weighted F1 scores for secondary structure prediction.}
    \begin{adjustwidth}{-.5in}{-.5in} 
    \label{tab:ss_f1_weight_ablation}
    \vskip 0.15in
    \begin{center}
        \begin{small}
            \begin{sc}
                \begin{tabular}{lcccccccc}
                    \toprule
                     & \multicolumn{2}{c}{RiNALMo} && \multicolumn{2}{c}{RNAstructure} && \multicolumn{2}{c}{Sequence Length} \\
                     \cline{2-3} \cline{5-6} \cline{8-9} \\
                    [-0.7em]
                    Test Family & F1\textsubscript{weighted}& F1\textsubscript{non-weighted} && F1\textsubscript{weighted} & F1\textsubscript{non-weighted} && Mean & STD \\
                    \midrule
                    5S rRNA & \textbf{0.884}  & \textbf{0.884} && 0.630 & 0.628 && 118.71 & 3.49 \\
                    SRP RNA & 0.628  & \textbf{0.700} && \textbf{0.631} & 0.633 && 180.23 & 100.81 \\
                    tRNA & \textbf{0.932}  & \textbf{0.931} && 0.703 & 0.704 && 77.10 & 5.02\\
                    tmRNA & \textbf{0.802}  & \textbf{0.801} && 0.431 & 0.433 && 366.02 & 27.72 \\
                    RNase P RNA & \textbf{0.803}  & \textbf{0.798} && 0.555 & 0.552 && 332.16 & 49.53 \\    
                    Group I intron & \textbf{0.675}  & \textbf{0.657} && 0.542 & 0.535 && 374.59 & 88.58 \\
                    16S rRNA & \textbf{0.721} & \textbf{0.736} && 0.525 & 0.570 && 317.04 & 145.61 \\
                    Telomerase RNA & 0.119 & 0.120 && \textbf{0.498} & \textbf{0.500} && 438.43 & 28.01 \\
                    23S rRNA & \textbf{0.840} & \textbf{0.848} && 0.716 & 0.730 && 326.20 & 51.95 \\
                    \midrule
                    Mean & \textbf{0.712} & \textbf{0.719} && 0.581 & 0.587 && 281.16 & 55.64\\
                    \bottomrule
                \end{tabular}
            \end{sc}
        \end{small}
    \end{center}
    \vskip -0.1in
    \end{adjustwidth}
\end{table*}

\section{Ablation Study}
\label{supp:ablation}
Here, we provide ablation study results evaluated on the downstream tasks. We study the effect of a few design and learning strategy choices. First, we compare the $650$M parameter RiNALMo with RiNALMo-$150$M and RiNALMo-$33$M and RNA-FM. Notice that RNA-FM \cite{chen2022_rna_fm} is a $100$M parameter LM with $640$ embedding dimensionality and $12$ layers. RiNALMo-$150$M and RiNALMo-$33$M were fine-tuned on the training datasets the same way as RiNALMo on all the downstream tasks. Second, we want to assess the influence of fine-tuning on the performance of RiNALMo on downstream tasks. To that end, we evaluate pre-trained and frozen RiNALMo, denoted as RiNALMo-frozen, on the downstream tasks. We utilize the output embeddings of RiNALMo-frozen as an input to the prediction heads for the downstream tasks. We only update prediction head weights during the training.

\begin{table*}[t]
    \caption{Inter-family generalization evaluation F1 scores for secondary structure prediction.}
    \label{tab:ss_ablation_app}
    \begin{adjustwidth}{-.5in}{-.5in} 
    \vskip 0.15in
    \begin{center}
        \begin{small}
            \begin{sc}
                \begin{tabular}{lcccccc}
                    \toprule
                    Test Family & RiNALMo & RiNALMo-frozen & RiNALMo-$150$M  & RiNALMo-$33$M & RNA-FM\\
                    \midrule
                    5S rRNA & \textbf{0.88}  & 0.86 & 0.85  & 0.55  & 0.57\\
                    SRP RNA & \textbf{0.70}  & \textbf{0.70} & 0.58  & 0.21 & 0.25 \\
                    tRNA & 0.93  & \textbf{0.95} & 0.91  & 0.85 & 0.79\\
                    tmRNA & 0.80  & 0.79 & \textbf{0.81}  & 0.43 & 0.28\\
                    RNase P RNA & \textbf{0.80}  & 0.77 & 0.77  & 0.49 & 0.31\\    
                    Group I intron & 0.66  & \textbf{0.67} & 0.39  & 0.27 & 0.16\\
                    16S rRNA & \textbf{0.74} & 0.68 & 0.72  & 0.43 & 0.14 \\
                    Telomerase RNA & \textbf{0.12} & \textbf{0.12} & 0.08  & 0.11 & 0.07\\
                    23S rRNA & 0.85 & 0.72 & \textbf{0.86}  & 0.56 & 0.19 \\
                    \midrule
                    Mean & \textbf{0.71} & 0.70 & 0.66 & 0.43 & 0.31\\
                    \bottomrule
                \end{tabular}
            \end{sc}
        \end{small}
    \end{center}
    \end{adjustwidth}
    \vskip -0.1in
\end{table*}

\begin{table*}[t]
    \begin{center}
        \caption{Performance of different models on SPOT-RNA's test dataset (TS0).}
        \label{tab:sec_struct_spot_rna_ablation}
        \vskip 0.15in
        \begin{small}
            \begin{sc}
                \begin{tabular}{lccc}
                    \toprule
                    Model & Precision & Recall & F1 \\
                    \midrule
                    RiNALMo & \textbf{0.784}  & \textbf{0.730} & \textbf{0.747} \\
                    RiNALMo-frozen & 0.663  & 0.663 & 0.652\\
                    RiNALMo-150M & 0.755  & 0.711 & 0.723\\
                    RiNALMo-$33$M & 0.718  & 0.673 & 0.685 \\
                    RNA-FM & 0.709  & 0.664 & 0.676 \\
                    \bottomrule
                \end{tabular}
            \end{sc}
        \end{small}
    \end{center}
    \vskip -0.1in
\end{table*}

\begin{table*}[t]
    \caption{Classification F1 score for multi-species splice-site prediction. We report the average value of donor and acceptor prediction results.}
    \label{tab:splice_site_ablation}
    \vskip 0.15in
    \begin{center}
        \begin{small}
            \begin{sc}
                \begin{tabular}{lcccc}
                    \toprule
                    Model & Fish & Fly & Plant & Worm  \\
                    \midrule
                    RiNALMo             & \textbf{97.47} & \textbf{95.83} & \textbf{95.38} & \textbf{94.86} \\
                    RiNALMo-frozen      & 81.02 & 68.18 & 67.73 & 63.75 \\
                    RiNALMo-$150$M      & 95.96 & 93.80 & 93.35 & 92.81 \\
                    RiNALMo-$33$M       & 94.86 & 93.08 & 90.83 & 91.23 \\
                    RNA-FM              & 94.94 & 93.21 & 89.93 & 90.79 \\
                    \bottomrule
                \end{tabular}
            \end{sc}
        \end{small}
    \end{center}
    \vskip -0.1in
\end{table*}

\begin{table*}[t]
    \caption{$R^2$ score and mean absolute error (MAE) for mean ribosome loading prediction for Random7600 and Human7600 datasets.}
    \label{tab:mrl_r2_ablation}
    \vskip 0.15in
    \begin{center}
        \begin{small}
            \begin{sc}
                \begin{tabular}{lccccc}
                    \toprule
                     & \multicolumn{2}{c}{Random7600} & & \multicolumn{2}{c}{Human7600} \\
                    \cline{2-3} \cline{5-6} \\
                    [-0.7em]
                    Model      & $R^2\uparrow$ & MAE $\downarrow$ & & $R^2\uparrow$ & MAE $\downarrow$ \\
                    \midrule
                    RiNALMo         & \textbf{0.925} & \textbf{0.282} & & \textbf{0.860} & \textbf{0.324}  \\
                    RiNALMo-frozen  & 0.610 & 0.626 & & 0.675 & 0.487 \\    
                    RiNALMo-$150$M  & 0.908 & 0.315 & & 0.836 & 0.351 \\
                    RiNALMo-$33$M   & 0.884 & 0.348 & & 0.827 & 0.359 \\
                    RNA-FM          & 0.842 & 0.404 & & 0.790 & 0.396 \\
                    \bottomrule
                \end{tabular}
            \end{sc}
        \end{small}
    \end{center}
    \vskip -0.1in
\end{table*}

The ablation study results evaluated on the secondary structure prediction tasks are given in Table~\ref{tab:ss_ablation_app} and Table~\ref{tab:sec_struct_spot_rna_ablation}, on the multi-species splice-site prediction task in Table \ref{tab:splice_site_ablation}, and on the mean ribosome load prediction task in Table \ref{tab:mrl_r2_ablation}. It can be seen that for all of the downstream tasks RiNALMo outperforms RiNALMo-$150$M, RNA-FM and RiNALMo-$33$M. This backs the usage of the larger RiNALMo model since it generally outperforms smaller models. Additionally, one can see that RiNALMo-$33$M performs comparably or sometimes even slightly better than RNA-FM despite a three times smaller model size. This indicates that the reason why RiNALMo outperforms RNA-FM is not just the model's scale but its novel architectural components as well. Unlike RNA-FM, RiNALMo uses more sophisticated Transformer architecture and more advanced techniques such as the RoPE and the SwiGLU activation function.

RiNALMo-frozen used the same prediction head as the other models in the downstream tasks but is the only one that was not fine-tuned. In Table~\ref{tab:ss_ablation_app}, it can be seen that most of the time RiNALMo-frozen is comparable to RiNALMo on the inter-family secondary structure prediction task, but is highly outperformed by the fine-tuned model for 23S rRNA and 16S rRNA families. Furthermore, RiNALMo-frozen outperforms RNA-FM on the inter-family secondary structure prediction task by a high margin. We can conclude that pre-trained RiNALMo captures much more structural information and knowledge than concurrent RNA-FM since it already performs comparable to the fine-tuned RiNALMo model and RNA-FM is nowhere near. We can relate the generalization capability of LMs for the inter-family secondary structure prediction tasks to the quality of pre-training. However, we see that fine-tuning can still lift the performance on the inter-family secondary structure prediction task. When tested on the intra-family secondary structure prediction task (Table \ref{tab:sec_struct_spot_rna_ablation}) and function prediction tasks (Tables \ref{tab:splice_site_ablation} and \ref{tab:mrl_r2_ablation}), it can be seen that the fine-tuned models outperform RiNALMo-frozen. There are two reasons for that. First, in the intra-family secondary structure prediction task, the model's generalization capability is not that important since the same RNA families can be seen in both training and test datasets. Second, RiNALMo is pre-trained on ncRNAs and the function prediction tasks are related to mRNAs previously not seen in the pre-training dataset. Fine-tuned models can better capture relations within RNA sequences not seen in pre-training.

Furthermore, we study the effect of architectural design choices. Our LM is advanced by modern architectural techniques such as rotary positional embedding (RoPE) \cite{su2024_roformer}, SwiGLU activation function \cite{shazeer2020_swiglu} and FlashAttention-2 \cite{dao2023_flashattention2}. In the ablation study, we start with a base model which is an RNA-FM-like model with sinusoidal positional embeddings and the GELU activation function. We assess the choice of replacing sinusoidal positional embeddings with RoPE in the first step and then replacing GELU with SwiGLU, resulting in our final RiNALMo architecture. All of the models in this ablation study employed FlashAttention-2, which is an IO-aware memory-efficient exact attention that does not affect performance. We used RiNALMo-$33$M in the ablation study because of the efficiency reasons and computational costs.

The ablation study results evaluated on the secondary structure prediction tasks are given in Table~\ref{tab:ss_ablation_33} and Table~\ref{tab:sec_struct_spot_rna_ablation_33}. In the tables, one can see that employing RoPE and SwiGLU instead of sinusoidal positional embeddings and GELU gradually boosts the performance of the LM on the secondary structure prediction tasks. This study supports our architectural design choices.

\begin{table*}[t]
    \caption{RiNALMo-$33$M ablation study evaluated on the inter-family secondary structure dataset.}
    \label{tab:ss_ablation_33}
    \vskip 0.15in
    \begin{center}
        \begin{small}
            \begin{sc}
                \begin{tabular}{lccc}
                    \toprule
                    Test Family & Base & Base + RoPE & Base + RoPE + SwiGLU \\
                    \midrule
                    5S rRNA & 0.27 & 0.37 & \textbf{0.55} \\
                    SRP RNA & 0.18 & 0.14 & \textbf{0.21} \\
                    tRNA & 0.52  & 0.59 & \textbf{0.85} \\
                    tmRNA & 0.26  & 0.31 & \textbf{0.43} \\
                    RNase P RNA & 0.21 & 0.41 & \textbf{0.49} \\    
                    Group I intron & 0.24 & 0.24 & \textbf{0.27} \\
                    16S rRNA & 0.21 & 0.37 & \textbf{0.43} \\
                    Telomerase RNA & 0.05 & 0.07 & \textbf{0.11}  \\
                    23S rRNA & 0.19 & 0.38 & \textbf{0.56} \\
                    \midrule
                    Mean & 0.24 & 0.32 & \textbf{0.43} \\
                    \bottomrule
                \end{tabular}
            \end{sc}
        \end{small}
    \end{center}
    \vskip -0.1in
\end{table*}

\begin{table*}[t]
    \caption{RiNALMo-$33$M ablation study evaluated on SPOT-RNA's test dataset (TS0).}
    \label{tab:sec_struct_spot_rna_ablation_33}
    \vskip 0.15in
    \begin{center}
        \begin{small}
            \begin{sc}
                \begin{tabular}{lccc}
                    \toprule
                    RiNALMo-$33$M & Precision & Recall & F1 \\
                    \midrule
                    Base & 0.706  & 0.667 & 0.676 \\
                    Base + RoPE & 0.704  & \textbf{0.682} & 0.682 \\
                    Base + RoPE + SwiGLU & \textbf{0.718}  & 0.673 & \textbf{0.685} \\
                    \bottomrule
                \end{tabular}
            \end{sc}
        \end{small}
    \end{center}
    \vskip -0.1in
\end{table*}

\backmatter
\bibliography{supplementary}